\documentclass[12pt]{article} 
\usepackage[koi8-r]{inputenc}
\usepackage[russian]{babel}
\usepackage{epsf}
\usepackage{epsfig}
\usepackage{amsmath}
\usepackage{color}
\usepackage{pazh}
\usepackage{natbib}
\tightenlines

\def\*{$^{*}$}
\def\а{$^{\mbox{\small а}}$}
\def\б{$^{\mbox{\small б}}$}
\def\в{$^{\mbox{\small в}}$}
\def\г{$^{\mbox{\small г}}$}
\def\д{$^{\mbox{\small д}}$}
\def\ергс{эрг~с$^{-1}$}
\def\ергсм{эрг~см$^{-2}$~с$^{-1}$}
\def\etal{{et~al.}}
\sloppypar

\begin{document}
\renewcommand{\figurename}{Fig.}
\renewcommand{\refname}{Bibliography}


\title{\bf Performance Analysis of Differential Speckle Polarimetry}

\author{\bf \copyright\, 2013 \ \ 
B. S. Safonov\affilmark{1*}}

\affil{
{\it Sternberg Astronomical Institute of Lomonosov Moscow State University, \\ Universitetskii pr-t, 13, Moscow, Russia}$^1$\\ 
}

\vspace{2mm}
\accepted{23 October 2012}

\sloppypar 
\vspace{2mm}
\noindent
We consider a method for obtaining information on polarization of astronomical objects radiation at diffraction limited resolution --- differential speckle polarimetry. As an observable we propose to use averaged cross spectrum of two short-exposure images corresponding to orthogonal polarizations, normalized by averaged power spectrum of one of images. Information on polarization can be extracted if object under study can be described by model with several parameters. We consider two examples: point-like source whose photocenter position depends on orientation of passing polarization and exozodiacal dust disc around a star. In first case the difference between photocenter positions can be measured with precision of 8~$\mu$as for 2.5-m telescope and 1.2~$\mu$as for 6-m telescope for object $V=13^m$. For second example method allows detection of discs around central star of $V=1^m$ with fractional luminosities of $1.8\times10^{-5}$ and $5.6\times10^{-6}$ for 2.5-m and 6-m telescope, respectively.



\noindent
{\bf keywords:\/} speckle interferometry, polarimetry


\vfill
\noindent\rule{8cm}{1pt}\\
{$^*$ E-mail: $<$safonov@sai.msu.ru$>$}

\clearpage

\section*{Introduction}

Many physical processes give rise to polarization of radiation emitted by astronomical objects and measurement of polarization is a powerful observational method \citep{TinbergenBook}. In visible light application of polarimetry in combination with high angular resolution is promising for study of the following astronomical objects: 1) circumstellar environment, 2) Solar System bodies, 3) active galactic nuclei. 


Here we consider a method for obtaining information on polarization of astronomical objects radiation at diffraction limited resolution --- differential speckle polarimetry (DSP). As follows from its title, this method represents a synthesis of speckle interferometry and differential polarimetry and presumes the use of instrument combining features of speckle interferometer and dual-beam polarimeter.


Combination of speckle interferometry and polarimetry has been implemented before by \citet{Falcke1996}. Their camera took images in different orientations of polarization serially, therefore the data were inevitably affected by differential errors. Nevertheless, authors were able to obtain new interesting information on polarization of optical radiation of $\eta$ Carinae with high angular resolution.


\citet{Schertl2000} constructed dual-beam polarimeter functioning as speckle interferometer. A term ``speckle polarimetry'', which we use here, was coined by authors of this work. The instrument was installed on 6-m telescope BTA of Special Astrophysical Observatory and worked in NIR band $K$. Images for each orientation of polarization were reconstructed separately by bispectrum \citep{Lohmann1983} and were processed by differential polarimetry techniques. These observations should demonstrate that polarized flux estimation had a greater precision than total flux estimation, but authors didn't indicated it.


The algorithm of DSP is based on generally accepted model of formation of instantaneous image in a focal plane of a telescope:
\begin{equation}
G(\boldsymbol{\alpha}) = O(\boldsymbol{\alpha}) \otimes T(\boldsymbol{\alpha}),
\end{equation}
where $G(\boldsymbol{\alpha})$ and $O(\boldsymbol{\alpha})$ --- distributions of light intensity in the focal plane and in the sky, respectively, $T(\boldsymbol{\alpha})$ --- instantaneous point spread function (PSF), $\boldsymbol{\alpha}$ --- vector of angular coordinate in the sky. For convenience let us assume that intensity distribution of object is normalized by total flux.


In Fourier space this equation becomes
\begin{equation}
\widetilde{G}(\boldsymbol{f}) = \widetilde{O}(\boldsymbol{f})\,\widetilde{T}(\boldsymbol{f}),
\label{eq:im_spec}
\end{equation} 
where $\widetilde{G}(\boldsymbol{f})$ --- Fourier spectrum of intensity distribution in the focal plane, $\widetilde{O}(\boldsymbol{f})$ --- visibility function of the object, $\widetilde{T}(\boldsymbol{f})$ --- instantaneous optical transfer function (OTF) of optical system ``atmosphere + telescope'', $\boldsymbol{f}$ --- vector of spatial frequency. OTF fluctuates over time due to perturbation of initially flat wavefront by atmospheric turbulence \citep{GoodmanBook}. As can be seen from the equation, these fluctuations are the source of multiplicative noise in $\widetilde{G}(\boldsymbol{f})$, which we will call ``atmospheric noise''.


\citet{Petrov1986} have shown that the Fourier spectrum $\widetilde{F}(\boldsymbol{f})$ of image received by the detector and normalised by the mean number of photon events $K$ forming it relates to the Fourier spectrum of intensity distribution if focal plane in a following way:
\begin{equation}
\widetilde{F}(\boldsymbol{f}) = \widetilde{G}(\boldsymbol{f}) + \eta(\boldsymbol{f}),
\label{eq:phot_add}
\end{equation}
where $\eta(\boldsymbol{f})$ --- spectrum of Poisson noise induced by quantum nature of light, it has circular symmetric complex normal distribution with variance equal to $K^{-1}$ for all frequencies.


If the optical system of telescope contains a beam-splitting polarizer, dividing light into two orthogonally polarized beams (e.g. Wollaston prism), it will form two images in focal plane, corresponding to orthogonal orientations of polarization. In other words, the system becomes a dual beam polarimeter. Let us assume that the first beam is horizontally polarized (subscript $h$) and the second is vertically polarized (subscript $v$). The equation (\ref{eq:im_spec}) for each of these images becomes:
\begin{equation}
\widetilde{G}_h(\boldsymbol{f}) = \widetilde{O}_h(\boldsymbol{f})\,\widetilde{T}_h(\boldsymbol{f}),~~~~\widetilde{G}_v(\boldsymbol{f}) = \widetilde{O}_v(\boldsymbol{f})\,\widetilde{T}_v(\boldsymbol{f}).
\label{eq:beams_spec2}
\end{equation} 
After substitution of these equations into (\ref{eq:phot_add}) we obtain:
\begin{equation}
\widetilde{F}_h(\boldsymbol{f}) = \bigl(\widetilde{O}_h(\boldsymbol{f}) \widetilde{T}_h(\boldsymbol{f}) + \eta_h(\boldsymbol{f})\bigr) e^{i\pi(\boldsymbol{\theta}_h\cdot\boldsymbol{f})},~~~~\widetilde{F}_v(\boldsymbol{f}) = \bigl(\widetilde{O}_v(\boldsymbol{f}) \widetilde{T}_v(\boldsymbol{f}) + \eta_v(\boldsymbol{f})\bigr) e^{i\pi(\boldsymbol{\theta}_v\cdot\boldsymbol{f})}.
\label{eq:beams_spec}
\end{equation}
Additional phase factors $e^{i\pi(\boldsymbol{\theta}_h\cdot\boldsymbol{f})}$ and $e^{i\pi(\boldsymbol{\theta}_v\cdot\boldsymbol{f})}$ are responsible for displacement of images by angles $\boldsymbol{\theta}_h$ and $\boldsymbol{\theta}_v$ in focal plane, induced by Wollaston prism. Taking account of these factors is necessary because in real experiment Fourier spectra of both images will be computed in the same coordinate system associated with the detector.

In speckle interferometry usually a large amount of frames is obtained. After that these frames are processed jointly. We also suppose that we have $M$ measurements of $\widetilde{F}_h(\boldsymbol{f})$ and $\widetilde{F}_v(\boldsymbol{f})$. Let us consider the following value:
\begin{equation}
\mathcal{R}(\boldsymbol{f}) = \frac{\langle \widetilde{F}_h(\boldsymbol{f}) \widetilde{F}_v^*(\boldsymbol{f}) \rangle_M}{\langle \widetilde{F}_v(\boldsymbol{f}) \widetilde{F}_v^*(\boldsymbol{f}) \rangle_M - K_v^{-1}},
\label{frac_def}
\end{equation}
where $\langle\dots\rangle_M$ means averaging over $M$ measurements. Value $\langle \widetilde{F}_h(\boldsymbol{f}) \widetilde{F}_v^*(\boldsymbol{f}) \rangle_M$ is an averaged cross spectrum of images\footnote{\citet{Petrov1986} considered a similar value, but defined for two spectral bands.}. Expression $\langle \widetilde{F}_v(\boldsymbol{f}) \widetilde{F}_v^*(\boldsymbol{f}) \rangle_M$ in denominator is an averaged power spectrum of image and represents biased estimation of averaged power spectrum of intensity distribution $\widetilde{G}_v(\boldsymbol{\alpha})$. Its bias equals $K_v^{-1}$ \citep{GoodmanBook}, therefore denominator of (\ref{frac_def}) is unbiased estimation of averaged power spectrum of intensity distribution. For $M>10$ denominator is significantly greater than zero given that the object visibility $\widetilde{O}_v$ is also different from zero. In this case expression (\ref{frac_def}) don't go into infinity.

Let us assume that the telescope is ideal $\widetilde{T}_h(\boldsymbol{f}) = \widetilde{T}_v(\boldsymbol{f})$ and the object is very bright $K_v~\ll~\langle \widetilde{F}_v \widetilde{F}_v^* \rangle_M$. Then, after substitution of (\ref{eq:beams_spec}) into (\ref{frac_def}) we get:
\begin{equation}
\mathcal{R}_0(\boldsymbol{f}) = \mathcal{R}(\boldsymbol{f}) e^{i\pi((\boldsymbol{\theta}_v-\boldsymbol{\theta}_h)\cdot\boldsymbol{f})}.
\label{frac_predef}
\end{equation}
One can see that except for the phase factor observable $\mathcal{R}$ can be considered as estimation of value $\mathcal{R}_0=\widetilde{O}_h / \widetilde{O}_v$, which depends on object properties only. Below we derive more strict formula for $\mathcal{R}$ and $\mathcal{R}_0$ accounting for the instrumental polarization and Poisson noise. We also propose a method of estimation of the phase factor.
%

Absolute value of $\mathcal{R}_0$ was used as an observable by Norris et al \citeyearpar{Norris2012}. They observed circumstellar envelopes of red supergiant stars by sparse aperture masking combined with dual-beam polarimetry on adaptive optics system NACO/VLT. Success of this work shows that parametric analysis of $\mathcal{R}_0$ can be quite fruitful. In contrast to that study we consider not only the absolute value of $\mathcal{R}_0$, but its argument also, we will call it ``phase'', by analogy with phase of the visibility function.


We extensively analysed the properties of $\mathcal{R}$, its dependence on instrumental polarization of optical system and its noise characteristics assuming that only Poisson and atmospheric noises are present. Unfortunately, in general case value $\mathcal{R}_0$ doesn't have any simple physical meaning. We propose to use simple parametric model fitting of $\mathcal{R}_0$ like it was done by Norris \etal \citeyearpar{Norris2012}. This approach is frequently applied in interferometry for analysis of visibility function measurements. For quantitative evaluations we use images simulated using Monte-Carlo method \citep{McGlamery1976,Harding1999}.



\section*{Theoretical analysis of $\mathcal{R}$ properties}
\label{sec:analysis}

Any real optical system possesses some instrumental polarization. In appendix A we consider how it affects the equation of image formation. There we have derived the Fourier spectra of intensity distributions in focal plane of dual-beam polarimeter corresponding to horizontal and vertical polarization (equations (\ref{eq:images1_A})) and expression for $\mathcal{R}$ (equation (\ref{eq:RM_gen0})).


From equation (\ref{eq:RM_gen0}) one can see that measured $\mathcal{R}$ depends on unknown phase factor $\mathcal{R} \sim e^{i\pi((\boldsymbol{\theta}_h-\boldsymbol{\theta}_v)\cdot\boldsymbol{f})}$. This factor can be measured by standard procedure, frequently used in differential polarimetry, --- exchange of images corresponding to horizontal and vertical polarization \citep{TinbergenBook}. This exchange can be performed by half-wave plate, installed before Wollaston prism. It allows to rotate the polarization axis of beam by $\pi/2$. The difference between corresponding measurement $\mathcal{R}^{\prime}$ from original measurement $\mathcal{R}$ is that phase factor enters it as $e^{i\pi((\boldsymbol{\theta}_v-\boldsymbol{\theta}_h)\cdot\boldsymbol{f})}$. Therefore it can be estimated from these two measurements by formula
\begin{equation}
e^{i\pi((\boldsymbol{\theta}_h-\boldsymbol{\theta}_v)\cdot\boldsymbol{f})} = \sqrt{\mathcal{R}(\boldsymbol{f})/\mathcal{R}^{\prime}(\boldsymbol{f})}.
\end{equation}
From now forth we assume that the phase factor is known and do not take account of it in derivation. Now we consider $\mathcal{R}$ as estimation of $\mathcal{R}_0$ and derive its two basic properties: bias and variance.


\subsection*{Bias of $\mathcal{R}$}
\label{subs:Rmean}

In appendix B it is demonstrated for mean of $\mathcal{R}(\boldsymbol{f})$ that
\begin{equation}
\overline{\mathcal{R}}(\boldsymbol{f}) = \mathcal{R}_0(\boldsymbol{f}) \bigl( 1 + \Delta \mathcal{R}(\boldsymbol{f}) \bigr).
\label{eq:RM_gen5}
\end{equation}
This equation says that measurement $\mathcal{R}(\boldsymbol{f})$ is biased relative to $\mathcal{R}_0(\boldsymbol{f})$. As follows from equation (\ref{eq:deltaR}), the value of bias $\Delta \mathcal{R}(\boldsymbol{f})$ depends on polarizing properties of telescope only.


\begin{figure}[h]
\begin{center}
\epsfxsize8.5cm
\epsffile{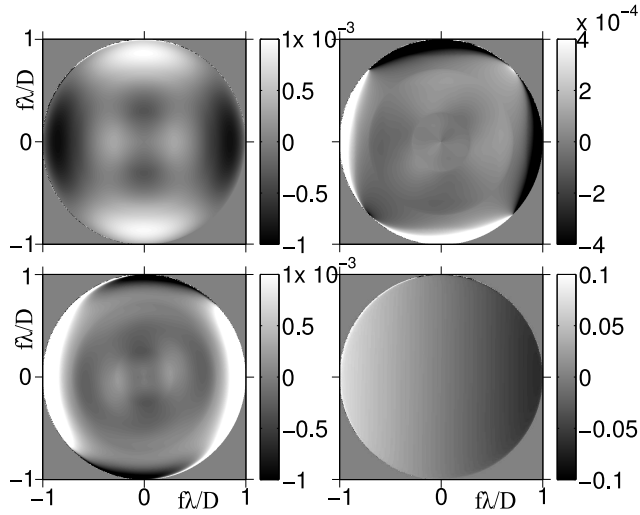}
\end{center}
\caption{Bias $\Delta \mathcal{R}(\boldsymbol{f})$ computed using formula (\ref{eq:deltaR}) from appendix B for Cassegrain (upper row) and Nasmyth (lower row) foci of the 2.5-m telescope. Left column --- real part of $\Delta \mathcal{R}(\boldsymbol{f})$, right column --- imaginary part of $\Delta \mathcal{R}(\boldsymbol{f})$. On the axes the normalized spatial frequency $f\lambda/D$ is plotted, where $\lambda$ --- wavelength, $d$ --- diameter of telescope. \label{fig:deltaR}}
\end{figure}

We have computed $\Delta \mathcal{R}(\boldsymbol{f})$ for Cassegrain and Nasmyth foci of the 2.5-m telescope, results are plotted in Fig.~\ref{fig:deltaR}. Real part of $\Delta \mathcal{R}(\boldsymbol{f})$ corresponds to amplitude of bias of $\mathcal{R}$ and has typical values of about $\pm5\times10^{-4}$ for both foci. The situation is different for imaginary part, which is responsible for the phase of bias of $\mathcal{R}$. For Cassegrain focus it varies by only $\approx10^{-4}$, but in case of Nasmyth it shows a significant overall slope with amplitude reaching 0.16. This slope is due to inclination of phase in Jones matrix of the telescope (see Fig.~\ref{fig:jones} in appendix A). Therefore we can argue that Cassegrain focus is more appropriate for precise measurements of $\mathcal{R}_0$.


Bias of $\Delta \mathcal{R}(\boldsymbol{f})$ can be measured with sufficient precision by means of observation of intentionally point-like and unpolarized star given that it and object of interest have similar spectra. From now on we assume that we have these measurements and $\mathcal{R}$ is an unbiased estimation of $\mathcal{R}_0$.


\subsection*{Variance of $\mathcal{R}$}
\label{Rnoise}

It is convenient to decompose the variance of $\mathcal{R}$ into variance of its amplitude $\sigma_{A}^2(\boldsymbol{f})$ and variance of its phase $\sigma_{\phi}^2(\boldsymbol{f})$. We estimated values of $\sigma_{A}^2(\boldsymbol{f})$ and $\sigma_{\phi}^2(\boldsymbol{f})$ using numerical simulation described in the next section. We generated 2000 pairs of images corresponding to horizontal and vertical polarization. Then for each image we computed Fourier spectrum and estimated variances $\sigma_{A}^2(\boldsymbol{f})$ and $\sigma_{\phi}^2(\boldsymbol{f})$ using formulae from appendix~C.

\begin{figure}[h]
\begin{center}
\epsfxsize=8.5cm
\epsffile{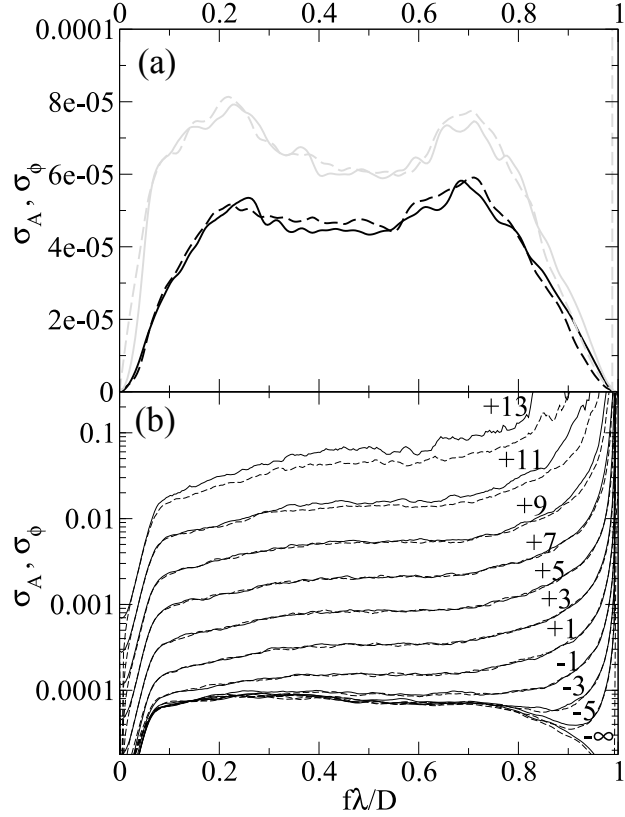}
\end{center}
\caption{Section of $\sigma_{A}$ (solid lines) and $\sigma_{\phi}$ (dashed lines), on the horizontal axis the normalized spatial frequency $f\lambda/D$ is plotted. {\it а} --- computation without atmospheric and with Poisson noises, black and grey lines --- for Cassegrain and Nasmyth foci, respectively; {\it b} --- computation with atmospheric and Poisson noises and for different magnitudes of object, values are given in figure. Computation details are given in text.\label{fig:wo_phaseV}}
\end{figure}

Let us consider the case of very bright object, in other words let us neglect Poisson noise. Results of corresponding computation are given in Fig.~\ref{fig:wo_phaseV}a as a section of standard deviation of $\sigma_{A}(\boldsymbol{f})$ and $\sigma_{\phi}(\boldsymbol{f})$ along X-direction for Cassegrain and Nasmyth foci. It is interesting that variances of amplitude and phase are nearly equal. Also it is noteworthy that variances for two foci differ only slightly, what is unexpected taking into account the fact that phase of Jones matrix fluctuates over the pupil much more greatly for Nasmyth focus than for Cassegrain (see Fig.~\ref{fig:jones} in appendix A). In order to explain this it should be recognised that noise of $\mathcal{R}$ is largely affected by difference of phases of Jones matrix elements $P_{A}$ and $P_{D}$. Meanwhile for Nasmyth focus most part of phase variation is caused by overall slope; this slope leads to bias of $\mathcal{R}$ (see previous section), but doesn't affect noise.


\begin{figure}[h]
\begin{center}
\epsfxsize=8.5cm
\epsffile{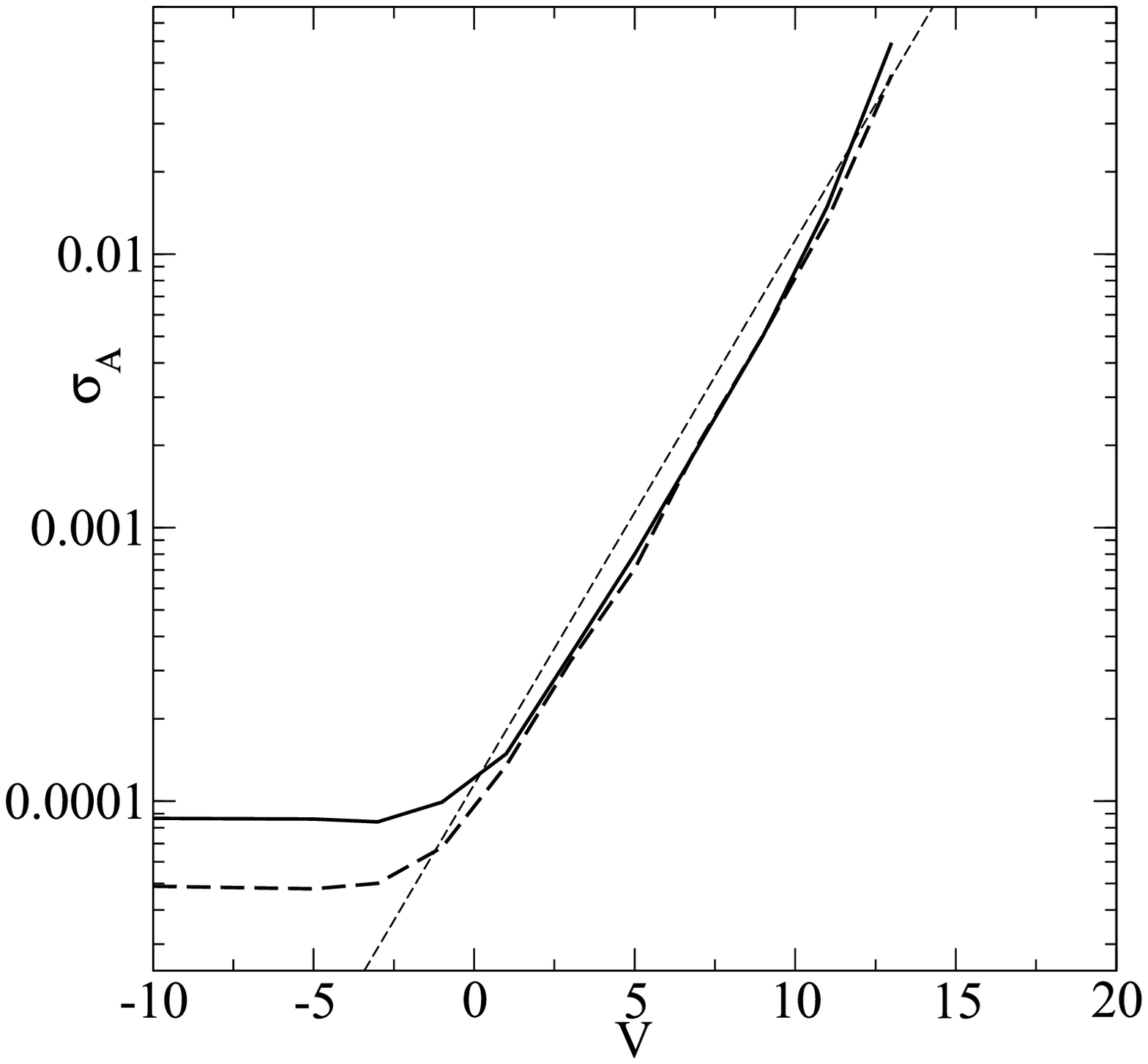}
\end{center}
\caption{Dependence of $\sigma_{A}$ on magnitude $V$ for $f = 0.4D/\lambda$, Nasmyth (thick solid line) and Cassegrain (thick dashed line) foci. The leftmost points are computed for very bright object. Computation details are given in the text. Thin dashed line stands for approximate estimation of Poisson noise computed using formula (\ref{eq:noisePetrov}).\label{fig:noise_mag}}
\end{figure}

In Fig.~\ref{fig:wo_phaseV}b we give the results of estimations of $\sigma_A(\boldsymbol{f})$ and $\sigma_{\phi}(\boldsymbol{f})$ accounting for the Poisson noise. These values rise at frequencies close to $D/\lambda$, this is caused by the fall of diffraction OTF ${\widetilde{T}}_0(\boldsymbol{f})$ at the same frequencies (formula (\ref{eq:noisePetrov})). The dependence of $\sigma_A(\boldsymbol{f})$ on magnitude can be seen more clearly in Fig.~\ref{fig:noise_mag} where it is plotted for spatial frequency $ 0.4D/\lambda$. One can see that Poisson noise dominates atmospheric noise for objects fainter than $V=-1^m$, i.e. for almost all astronomical objects. In domain of Poisson noise prevalence the approximate formula from \citep{Petrov1986} describes behaviour of variance reasonably well (thin dashed line in Fig.~\ref{fig:noise_mag}):
\begin{equation}
\sigma_{\phi}^2(\boldsymbol{f}) = {\Bigl( M {|\widetilde{I}(\boldsymbol{f})|}^2 {\widetilde{T}}_0(\boldsymbol{f}) K L^{-1}\Bigr)}^{-1},
\label{eq:noisePetrov}
\end{equation}
where $\widetilde{I}(\boldsymbol{f})$ --- visibility function of object (here we assumed that object appearance doesn't depend on polarization significantly), $L=2.3{(D/r_0)}^2$ --- mean number of speckles in the image, $r_0$ --- Fried parameter. In domain of Poisson noise prevalence difference between noises for Cassegrain and Nasmyth foci essentially disappears, consequently hereafter we consider the Nasmyth focus only.

\section*{Monte-Carlo simulation of DSP}
\label{subs:sim_params}

In order to estimate the expected performance of the DSP we simulated the process of measurements using Monte-Carlo method \citep{McGlamery1976,Harding1999}. This method is based on generation of a large number of instantaneous images, disturbed by the atmosphere, followed by simulation of subsequent processing.


We modelled the atmosphere as a set of infinitely thin turbulent layers. Each of these layers is defined by altitude, turbulence intensity, wind speed and direction. For each layer we generated a realisation of random phase screen\footnote{Phase screen is an idealized infinitely thin optical component, which affects only phase of passing wavefront.} constituting $N\times N$ matrix with von Karman power spectrum:
\begin{equation}
F_{\phi}(\boldsymbol{f})=0.0229\,r_0^{-5/3}{(f^2+L_0^{-2})}^{-11/6},
\label{eq:vonKarman}
\end{equation}
$f$ --- absolute value of spatial frequency vector $\boldsymbol{f}$, $r_0$ --- Fried parameter, corresponding to turbulence intensity of this layer, $L_0$ --- outer scale of turbulence, we adopted typical value of $L_0=25$~m.


Originally flat wavefront propagated these phase screens, starting with the highest, in output we obtained disturbed wavefront. Then we took into account that instrument is dual-beam polarimeter. In order to do this we made two copies of wavefront. Then in accordance with Jones matrix definition \citep{BornWolf} we multiplied one copy by $P_{A}$ and another by $P_{D}$, see appendix A.


Calculation of $P_{A}$ and $P_{D}$ was made by use of {\it Zemax} software for three optical systems: Cassegrain and Nasmyth foci of the future 2.5-m telescope of SAI and for primary focus of the BTA. For the 2.5-m telescope the central obscuration is 0.43, for BTA this parameter equals 0.33. We assumed that mirrors are coated with bare aluminium\footnote{Transparent protective coating of mirrors can affect the instrumental polarization greatly. Therefore for mirrors of real telescope instrumental polarization can be substantially different from our calculations.}. Amplitude and phase of calculated Jones matrix for the 2.5-m telescope are presented in Fig.~\ref{fig:jones} in appendix A.


After this we computed disturbed images and added the Poisson noise to them. We have taken into consideration two-fold multiplicative noise of EMCCD detector \citep{Hynecek2003}, optimal for speckle interferometry. Thus we have accounted for all effects of interest for our problem: atmospheric disturbance, instrumental polarization and Poisson noise.


Let us discuss some parameters of the simulation. Computation was performed for $V$ band, some parameters were also computed for $I$ band. Total efficiency in $V$ and $I$ was 0.54 and 0.43, respectively (with consideration of transmission of atmosphere, telescope and instrument optics, quantum efficiency of detector). We adopted value of 30~ms for exposure, what is typical for speckle interferometric observations. For objects fainter than $1^{m}$ we used minimal period of exposures 30~ms, as long as in this domain uncorrelated Poisson noise dominates. For bright objects period was increased to 120~ms for adequate account for atmospheric noise.


Uncorrelatedness of Poisson noise allowed us to compute quantitative performance characteristics --- variances of estimations --- for small number of frames, e.g. 100, and then adapt them for long series by multiplication by square root of frame number ratio. Standard series duration was adopted as 1 hour, what is much longer than typical series in speckle interferometry. For correct processing of speckle interferometric measurements atmospheric conditions (Speckle Transfer Function --- STF) should be constant throughout the series. For DSP there is no such requirement.


As a model of atmosphere we adopted two turbulent layers of equal intensity, yielding seeing 0.91$^{\prime\prime}$ in $V$ band. This is median seeing for Mt. Shatdzhatmaz, supposed location of the 2.5-m telescope \citep{Kornilov10}. Both layers move with speed of 10~m/s in opposite directions, what provides maximum temporal uncorrelatedness of atmospheric noise. This rough model is sufficient as long as we are not interested in temporal or isoplanatic characteristics of image. For convenience of comparison computation for BTA was performed for the same model.


Resulting series of images ${F}_{h}$ and ${F}_{v}$ were used for estimation of $\mathcal{R}$ by formula (\ref{frac_def}). Its variance was estimated by formulae from appendix C.



\section*{Parametric analysis of $\mathcal{R}$}
\label{parametric}

Fitting measurements of $\mathcal{R}$ with a model of a few parameters is the simplest way of extraction of information from it. We consider a special case of such a model, having one parameter:
\begin{equation} 
\Phi(\boldsymbol{f},p) = \Phi_0(\boldsymbol{f}) + p \Phi_1(\boldsymbol{f}).
\label{eq:lin_model}
\end{equation} 
In this case the problem resolves itself into finding of parameter $p$, for which function $\Phi(\boldsymbol{f},p)$ describes the observations in the best way. We estimate the parameter by minimization of sum of squared deviations of model from measurements, taken with weights inversely proportional to estimated variances of measurements.


Error of $p$ estimation, or equivalently its expected variance $\sigma_p^2$, defines the DSP efficiency. For given linear model it is derived by the following expression (\citet{Kuzmenkov1985} considered more general case):
\begin{equation}
\frac{1}{\sigma_p^2} = \sum_i \frac{\Phi_1^2(\boldsymbol{f}_i)}{\sigma_i^2},
\label{eq:par_sum}
\end{equation}
Summation is being performed over ${(f_c/f_0)}^{2}$ independent measurements, where $f_c=D/\lambda$ --- cutoff frequency, $\sigma_i^2$ --- variances of measurements, they can be estimated using formulae (\ref{eq:RnoiseA}) and (\ref{eq:RnoisePhi}). $\boldsymbol{f}_i$ --- coordinates of measurements of $\mathcal{R}$ in Fourier space.


Now we consider two parametric models: 1) point-like object, whose photocenter position depends on orientation of polarization; 2) point-like object and faint polarized envelope.


\subsection*{Point-like object, whose photocenter position depends on orientation of polarization}
\label{astrom}

Suppose that we have an object whose appearance depends on orientation of polarization $\widetilde{O}_h \ne \widetilde{O}_v$, therefore $\mathcal{R}_0 \ne 1$. Let us denote typical angular extent of this object as $\gamma$, what corresponds to spatial frequency domain $f_{\gamma}\approx1/\gamma$. For frequencies $|f| \ll f_{\gamma}$ amplitude and phase of $\mathcal{R}_0$ can be represented by Tailor series with small parameter $f/f_{\gamma}$. In this domain variation of amplitude is dominated by quadratic term and variation of phase --- by linear term. If object demonstrates high asymmetry of polarized and/or total flux then $\mathcal{R}_0$ phase changes significantly at frequencies $\approx f_{\gamma}$. In this case in the domain $|f| \ll f_{\gamma}$ variation of argument is much larger than variation of amplitude, moreover, most part of this variation consists in slope. Consequently, for $\mathcal{R}_0$ in domain $|f| \ll f_{\gamma}$ one can write:
\begin{equation}
\mathcal{R}_0 \approx \exp{\{i \pi (\boldsymbol{\theta} \boldsymbol{f})\}},
\end{equation}
where $\boldsymbol{\theta}$ --- vector of difference between photocenters of images corresponding to different orientations of polarization. In reality frequencies $|f| \ll f_{\gamma}$ are sampled when the object is much smaller than diffraction resolution of telescope. Thus $\boldsymbol{\theta}$ becomes the most robustly defined parameter of object.


For the described type of object it is convenient to model the phase of $\mathcal{R}$. In this case the model is $\Phi(\boldsymbol{f},\theta_x,\theta_y) = \pi (f_x\theta_x + f_y\theta_y)$, where $\theta_x,\theta_y$ --- components of vector of photocenter position difference. We take the $\theta_x$ as a sought quantity, $\theta_y$ is assumed to be known. This simplification is valid as long as mutual estimation of $\theta_x$ and $\theta_y$ is uncorrelated. In terms of the linear model (\ref{eq:lin_model}) $\Phi_1(\boldsymbol{f})=\pi f_x$. Taking this into consideration we can rewrite equation (\ref{eq:par_sum}) for the parameter $\theta_x$:
\begin{equation}
\frac{1}{\sigma_x^2} = \sum_i \frac{\pi^2 f_{xi}^2}{\sigma_{\phi i}^2},
\label{eq:par_integ_anoise}
\end{equation}
where $f_{xi}$ --- X-component of vector $\boldsymbol{f}_i$, phase noise $\sigma_{\phi i}^2 = \sigma_{\phi}^2(\boldsymbol{f}_i)$ is derived using method from appendix C.

\begin{figure}[h]
\begin{center}
\epsfxsize=8.5cm
\epsffile{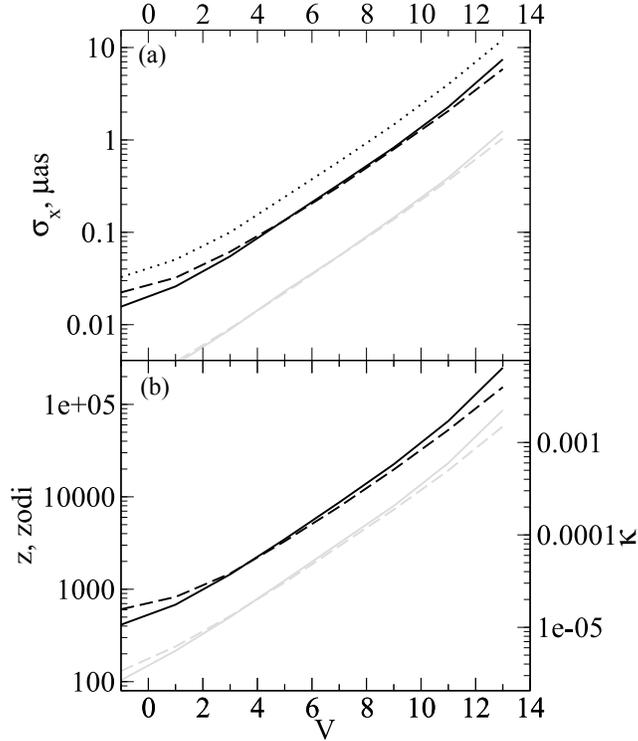}
\end{center}
\caption{{\it a} --- dependence of expected differential astrometry precision on magnitude $V$. {\it b} --- dependence of expected precision of $z$ estimation on magnitude $V$ for exozodiacal disc analogous to disc around $\tau$ Ceti. $z$ --- flux of disc relative to flux of solar zodiacal disc. Along right vertical axis the contrast $\kappa$ of disc is plotted. For both figures: solid lines --- $r_0$=12.5~cm (median conditions at Mt. Shatdzhatmaz), dashed lines --- $r_0=20.5$ (10\% of best conditions at Mt. Shatdzhatmaz). Black lines --- 2.5-m telescope, grey lines --- 6-m telescope. Dotted line stands for median conditions, 2.5-m telescope and $I$ band.
\label{fig:par_noise}}
\end{figure}

Using this expression we computed expected variance of angle $\theta_x$ for adopted conditions of simulation and plotted it in Fig.~\ref{fig:par_noise}a. It can be seen that expected variance doesn't depend on seeing. This feature of differential speckle interferometry was noted earlier by \citep{Petrov1986}.


The figure shows that for object $V=13^m$, precision of estimation of difference between photocenters corresponding to horizontal and vertical polarizations is $8~\mu$as for the 2.5-m telescope and $1.2~\mu$as for the 6-m telescope. Such precision significantly exceeds typical precision of differential narrow-field astrometry at 5-8~m telescopes, which reaches 100~$\mu$as and 150-200~$\mu$as with and without adaptive optics, respectively \citep{Cameron2009,Pravdo1996,Lazorenko2006}.


The detection and measurement of differential astrometric signal can be particularly productive for the study of BL~Lac objects. These active galactic nuclei (AGN) periodically show increase in polarization degree of 10-20\% on timescales of 50-100 days. It would be interesting to measure the polarized flux photocenter movement accompanying these events.


\subsection*{Exozodiacal disc}
\label{exozodi}

Various processes associated with stellar evolution frequently give rise to the occurrence of circumstellar matter, e.g. dust disc. Radiation of dust has two main components. The first is thermal IR radiation of dust heated by nearby star. At the moment observations and analysis of IR radiation of dust is the main source of information about its distribution in circumstellar space, temperature and chemical composition. This is caused by the fact that in IR the contrast between disc and star is not so large as in visible \citep{diFolco2007}. The second component of dust radiation is scattered visible light of star. Imaging of dust discs in scattered light, especially in multiple photometric bands is a powerful tool of dust diagnosis, which supports IR observations \citep{Absil2010}. Main difficulty of circumstellar dust observations is high contrast between star and envelope --- $10^{-5}$ and higher. However scattered light is polarized what simplifies the problem of its detection and characterization. In this subsection we determine limiting contrast between star and disc for DSP using numerical simulation.

\begin{figure}[h]
\begin{center}
\epsfxsize=8.5cm
\epsffile{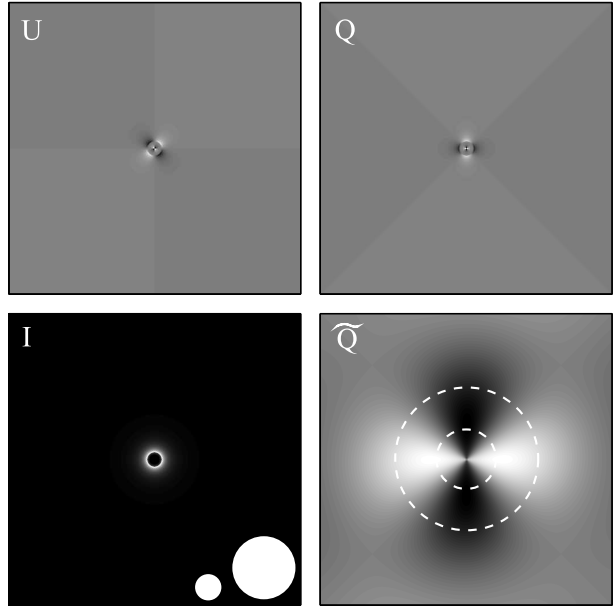}
\end{center}
\caption{The model of exozodiacal disc around $\tau$ Ceti, computed using ZODIPIC (for details see the text). Upper row: Stokes parameters $U$ and $Q$, lower row, left panel --- total intensity $I$, circles diameters correspond to $1.22\lambda/D$ for $D=2.5$~m (large circle) and $D=6$~m (small circle). Lower right panel --- Fourier spectrum $\widetilde{Q}_d$, circles stand for domains available for measurements with the 2.5-m (small circle) and the 6-m (large circle) telescope.\label{fig:zodiStokes}}
\end{figure}

Let us assume the following simple model of object: exozodiacal disc, analogous to solar one, but with total luminosity $z$ times higher, rotating around $\tau$~Ceti ($L = 0.52 L_{\odot}$, distance 3.65~pc, magnitude $V=3.5^m$), visible from the pole. Parameter $z$ is a measure of disc's ``magnitude'' relative to solar one, a term ``zodi'' was adopted in literature as a unit for it \citep{Roberge2012}. The disc was simulated with ZODIPIC \citep{Moran2004ZODIPIC} package for IDL. This program makes use of data on the Solar zodiacal disc from \citep{Moran2004ZODIPIC}. Scattering by dust is modeled in accordance with paper \citep{Hong1985}. We augmented the code with the computation of Stokes vector components $Q_{d}(\boldsymbol{\alpha})$ and $U_{d}(\boldsymbol{\alpha})$, corresponding equations also were taken from \citep{Hong1985}. Obtained distributions of Stokes parameters $I$, $Q$, $U$ are given in Fig.~\ref{fig:zodiStokes}. 


Visibilities $\widetilde{O}_h$ and $\widetilde{O}_v$ for this type of objects are
\begin{equation}
\widetilde{O}_h = \widetilde{I}_{star} + z\widetilde{I}_d + z\widetilde{Q}_d, ~~~~ \widetilde{O}_v = \widetilde{I}_{star} + z\widetilde{I}_d - z\widetilde{Q}_d,
\end{equation}
where $\widetilde{I}_{star}$ --- visibility function of the star, $\widetilde{I}_d$ and $\widetilde{Q}_d$ --- Fourier spectra of Stokes vector distribution of the disc. The $\widetilde{Q}_d$ for our model of disc is also plotted in Fig.~\ref{fig:zodiStokes}. In spite of the fact that disc angular size is formally below diffraction limit of the 2.5-m telescope, evidence of its presence in $\widetilde{Q}_d$ extends into Fourier space domain available for measurements with the 2.5-m telescope (i.e. at frequencies smaller than cutoff frequency). The possibility to obtain information about source features smaller than diffraction resolution of telescope is known as super-resolution \citep{MartiVidal2012}.


Value $\mathcal{R}_0$ for this model becomes
\begin{equation}
\mathcal{R}_0 = \frac{\widetilde{I}_{star} + z\widetilde{I}_d + z\widetilde{Q}_d}{\widetilde{I}_{star} + z\widetilde{I}_d - z\widetilde{Q}_d}.
\end{equation}


Let us assume that total luminosity of envelope is smaller than stellar luminosity $I_d \ll I_{star}$:
\begin{equation}
\mathcal{R}_0 \approx \biggl( 1 + \frac{z\widetilde{Q}_d}{\widetilde{I}_{star}} \biggr) \biggl( 1 + \frac{z\widetilde{Q}_d}{\widetilde{I}_{star}} \biggr) \approx 1 + \frac{2 z \widetilde{Q}_d}{\widetilde{I}_{star}}.
\label{simpleR}
\end{equation}
Now we take into account that star is point-like source and its visibility $\widetilde{I}_{star} = 1$:
\begin{equation}
\mathcal{R}_0 \approx 1 + 2z\widetilde{Q}_d.
\end{equation}
Thus parametric model, approximating observed $\mathcal{R}$, is:
\begin{equation}
\Phi(\boldsymbol{f},z) = 1 + 2z\widetilde{Q}_d(\boldsymbol{f}).
\label{eq:zodi_model}
\end{equation}
The aim of fitting is finding of $z$. Another consequence from this equation is that it is possible to obtain Fourier spectrum of polarized flux of circumstellar envelope and then its image. However image has a very large number of unknowns (pixels intensities), each of them will be estimated with low precision. That is why for faint discs model with small number of parameters (in our case just one) is more appropriate.


In this case we will be interested in amplitude of $\mathcal{R}$ only, because the source is centrosymmetrical, therefore phase of $\mathcal{R}$ equals zero. Equation (\ref{eq:par_sum}) for the error of $z$ will be:
\begin{equation}
\frac{1}{\sigma_z^2} = \sum_i \frac{4\widetilde{Q}_{di}^2}{\sigma_{Ai}^2},
\label{eq:par_integ_znoise}
\end{equation}
where $\widetilde{Q}_{di} = \widetilde{Q}_{d}(\boldsymbol{f}_i)$, noise of amplitude $\sigma_{Ai}^2 = \sigma_{A}^2(\boldsymbol{f}_i)$ is being estimated by formula (\ref{eq:RnoiseA}).


Parameter $z$ estimated from measurements is a normally distributed random number, therefore probability of detection of disc $z=\sigma_z$ is $\approx70\%$. Although this threshold corresponds to certain model, it provides a glimpse of method performance in observations of circumstellar environment.


Estimations of detection limit for the 2.5-m and the 6-m telescopes and different seeing conditions are presented in Fig.~\ref{fig:par_noise}b. One can see that in this case seeing also weakly affects performance.


At the 2.5-m telescope DSP will allow detection of discs $z=700$~zodi around $V=1^m$ stars, at 6-m telescope --- $z=220$~zodi (absolute contrast between the disc and the star $1.8\times10^{-5}$ and $5.7\times10^{-6}$, respectively). Exozodiacal disc around Vega is one of brightest discs, in IR its luminosity reaches 3000~zodi. In visible luminosity of the disc is likely to be lower \citep{Absil2006}\footnote{The data on exozodiacal discs luminosity in visible are quite scarce due to mentioned difficulties of high contrast.}. Thus we can conclude that brightest exozodiacal discs are available for study with DSP at the 2.5-m and the 6-m telescopes.


Exozodiacal discs are only special case of circumstellar envelopes, they are typical for relatively old main sequence stars. In other situations contrast between star and envelope is much lower. E.g. for envelope observed by \citep{Norris2012}, fraction of scattered light $\approx10\%$, therefore it can be detected by DSP easily. Young stars also frequently possess strong dust envelopes (see e.g. \citep{Krist2005}). These objects are of great interest and results of this subsection can be used for estimation of possibility of their observation with DSP.


\section*{Practical aspects of DSP}
\label{sec:disc}

Specifics of DSP impose some constraints on device implementing this method. Let us discuss the most important of them.


Fourier spectra used for calculation of $\mathcal{R}$ can be correctly estimated if pixel angular size is less than $\lambda/2D$. EMCCD detector is preferable as long as it has negligibly small readout noise and allows to make short exposures at sufficiently high frequency.


Theoretical analysis shows that $\mathcal{R}$ is biased estimation of value $\mathcal{R}_0=\widetilde{O}_h / \widetilde{O}_v$, which depends on object properties only. Bias of $\mathcal{R}$ depends on polarizing properties of telescope and instrument, for measurement of it we propose two stages of calibration.


The first stage is exchange of images corresponding to horizontal and vertical polarizations, what can be implemented with half-wave plate rotating polarization axis by $\pi/2$. This procedure will allow to precisely measure phase factor from equation (\ref{frac_predef}), or equally the difference between deviation angles of Wollaston prism. Differential aberrations occurring in optical system after half-wave plate also will be contained in this phase factor. Thus it makes sense to install half-wave plate as early as possible in optical system.


The second stage of calibration is required for removal of differential aberrations induced by optical elements before half-wave plate (e.g. telescope mirrors). This calibration consists in measurement of $\mathcal{R}$ of some reference star. This star should have similar spectrum and be brighter than object of interest by $4-5^m$ (when it is possible), otherwise calibration measurement will significantly increase the error of final estimation. Appearance of reference star shouldn't depend on orientation of polarization axis. This condition will be fulfilled more than enough for single main sequence star of spectral class later than A0 having angular size less than $\approx10$~mas.


Differential aberrations give rise to not only bias but atmospheric noise as well. Therefore they should be minimized in design phase. This can be achieved by ensuring that paths of beams corresponding to horizontal and vertical polarizations are as close as possible to each other. In order to do this Wollaston prism should be installed as last optical element in system. In this case the most part of differential aberrations will be induced by the prism itself due to birefringency of its material.


Deviation of any surface of the prism intersected by beams from the plane by $\Delta$ results in differential phase aberration $k(n_o-n_e)\Delta$, where $n_o$ and $n_e$ --- refractive indices corresponding to ordinary and extraordinary beams. Value $n_o-n_e$ for calcite is $-0.172$, for quartz $+0.009$. Thus the use of quartz prism is more appropriate, as long as surface quality requirements are relaxed for quartz in comparison with calcite. However deviation angle for quartz prism is significantly smaller.


Let us estimate acceptable level of differential aberrations. Differential astigmatism (similar to plotted in Fig.~\ref{fig:jones} in appendix A, lower right, phases of $P_A$ and $P_D$) $Z_5=0.011$, what corresponds to amplitude of phase variation $0.054$~rad, results in increase of atmospheric noise by factor of 3 in Cassegrain focus. Corresponding deviation of prism surface from plane is $\lambda/20$ for calcite and $\lambda$ for quartz. In this case atmospheric noise mainly affects observations of bright objects $V<4^m$, because for them Poisson noise is sufficiently small (see Fig.~\ref{fig:noise_mag}). If we restrict ourselves to faint objects, requirements for differential aberrations are substantially relaxed.


In analysis being performed in this work we take into account three effects: instrumental polarization of telescope, atmospheric disturbances and finiteness of number of detected photons. Measurements made with real instrument will probably be affected by other systematic errors induced by other factors. However some of these errors will likely to be eliminated by described calibrations.


Authors of \citep{Canovas2011} achieved significant success in removing of systematic errors in differential polarimetry with Extreme Polarimeter (ExPo) instrument at 4.2-m WHT telescope. This instrument is quite similar to one supposed by us, it obtains a large series of short-exposure images taken in orthogonal polarizations simultaneously. ExPo exchanges the images after every exposure for elimination of systematical errors. It is probable that this approach will be effective for DSP also.


\section*{Conclusions}
\label{sec:conc}

Differential polarimetric measurements with high angular resolution are much more precise than absolute what provides additional opportunities of study of astronomical objects. Improved precision arises out of the fact that many factors disturbing wavefront affect its polarization components equally. Atmospheric turbulence limiting angular resolution of ground based telescopes is one of these factors. DSP allows to significantly reduce atmospheric noise influence and obtain information about polarization properties of object's radiation with diffraction resolution.


As an observable for DSP we have adopted value $\mathcal{R}$ --- averaged cross spectrum of two short-exposure images corresponding to perpendicular orientations of polarization, normalized by averaged power spectrum of one of images. In order to analyse the properties of $\mathcal{R}$ noise we perform numerical simulation of DSP using Monte-Carlo method. The main outcome of this simulation is that Poisson noise dominates atmospheric noise for stars fainter than $V=-1^m$ given that there isn't any differential aberrations apart from induced by the telescope.


Measurements of $\mathcal{R}$ can be interpreted by parametric model fitting. Using numerical simulation we study the performance of method for two examples of such model. In first example we consider point-like object whose photocenter position depends on orientation of polarization. In this case difference between photocenter positions for object with $V=13^m$ can be measured with precision of $8~\mu$as and $1.2~\mu$as at the 2.5-m and the 6-m telescopes, respectively. Such astrometry precision would provide interesting opportunities of investigation of some astronomical objects, e.g. AGN. In second example we evaluate capabilities of detection and characterization of circumstellar dust discs using idealized model of such disc. It is found out that for $V=1^m$ star discs with luminosity of $1.8\times10^{-5}$ and $5.6\times10^{-6}$ relative to the star can be detected at the 2.5-m and the 6-m telescope, respectively. Such contrast is typical for brightest exozodiacal discs and dust envelopes around young stars. Performance of DSP depends on the seeing weakly.


It is interesting to compare capabilities of DSP with other existing and planned polarimeters. At the moment one of the pressing observational problems in astronomy is the detection of faint components (mainly exoplanets) in close vicinity of bright stars. There are several existing and planned instruments designed to solve this problem: HiCIAO/SUBARU \citep{Hashimoto2011}, NACO/VLT \citep{Norris2012}, SPHERE-IRDIS/VLT \citep{Dohlen2008}, SPHERE-ZIMPOL/VLT \citep{Thalmann2008}, GPI/Gemini \citep{Macintosh2006}, MMT-POL/MMT \citep{Packham2010b}. Some of them will be able to reach contrast of $10^{-8}$. All these  instruments are very complex and expensive devices having differential polarimetry mode and fed through adaptive optics. Meanwhile all of them except ZIMPOL are designed for NIR.


ExPo instrument works in visible, but its processing algorithm operates with images, therefore its angular resolution $0.5^{\prime\prime}$ is limited by the seeing. Angular size of pixel of this camera is larger than $\lambda/2D$, therefore the data obtained with it cannot be processed with DSP.


Thus it can be seen that DSP has its niche --- relatively high contrast even in visible at small angular separations from the star and high precision of differential astrometry.


Author is grateful to Tokovinin A.A. for discussion of speckle interferometry issues. Comments from Kornilov V.G. and anonymous referee helped to substantially improve the paper. Author acknowledges support from ``Dynasty'' foundation.




\section*{Appendix A. Polarization in focal plane of telescope}
\label{app:focal_pol}

Vector $\widetilde{\boldsymbol{S}}_{\!f(h,v)}(\boldsymbol{f})$ of Fourier spectra of Stokes parameters distribution in the focal plane of the telescope relates to similar vector $\widetilde{\boldsymbol{S}}(\boldsymbol{f})$ defined in the sky in the following way \citep{Almeida1992}:
\begin{equation}
\widetilde{\boldsymbol{S}}_{\!f(h,v)}(\boldsymbol{f}) = \widetilde{\boldsymbol{M}}_{\!f(h,v)}(\boldsymbol{f}) \widetilde{\boldsymbol{S}}(\boldsymbol{f}).
\label{eq:focal_Mueller_prod}
\end{equation}
This vector equation is written down for both images corresponding to horizontal ($h$) and vertical ($v$) polarizations. $\widetilde{\boldsymbol{M}}_{\!f(h,v)}(\boldsymbol{f})$ are Fourier spectra of Mueller matrices of the telescope. Equation (\ref{eq:focal_Mueller_prod}) is valid as long as $\widetilde{\boldsymbol{M}}_{\!f(h,v)}(\boldsymbol{f})$ doesn't depend on direction in the sky. This condition may be considered fulfilled because we deal with narrow fields $\approx2^{\prime\prime}$. Equation (\ref{eq:focal_Mueller_prod}) constitutes some generalization of the common equation of image formation in Fourier space, in this sense $\widetilde{\boldsymbol{M}}_{\!f(h,v)}(\boldsymbol{f})$ is generalized OTF.


The method of computation of matrix $\widetilde{\boldsymbol{M}}_{\!f(h,v)}(\boldsymbol{f})$ is given in \citep{Almeida1992} and \citep{Azzam1987}. Here we briefly reproduce it:
\begin{equation}
\widetilde{\boldsymbol{M}}_{\!f(h,v)}(\boldsymbol{f}) = \boldsymbol{B} \widetilde{\boldsymbol{T}}_{\!h,v}(\boldsymbol{f}) \boldsymbol{B}^{-1}, 
\label{eq:spec_Mueller}
\end{equation}
where matrix $\boldsymbol{B}$ is defined \citep{Azzam1987} as
\begin{equation}
\boldsymbol{B} = \begin{pmatrix}
1 & 0 & 0 & 1 \\
1 & 0 & 0 & -1 \\
0 & 1 & 1 & 0 \\
0 & i & -i & 0 \\
\end{pmatrix}.
\end{equation}


Matrices $\widetilde{\boldsymbol{T}}_{\!h}(\boldsymbol{f})$ and $\widetilde{\boldsymbol{T}}_{\!v}(\boldsymbol{f})$ also have size $4\times4$. Given that Wollaston prism is ideal all elements of $\widetilde{\boldsymbol{T}}_{\!h}(\boldsymbol{f})$ equal zero except for the elements of the first row: $\widetilde{T}_{AA^{*}}(\boldsymbol{f}), \widetilde{T}_{AB^{*}}(\boldsymbol{f}), \widetilde{T}_{BA^{*}}(\boldsymbol{f}), \widetilde{T}_{BB^{*}}(\boldsymbol{f})$. The same is valid for matrix $\widetilde{\boldsymbol{T}}_{\!v}(\boldsymbol{f})$, however in this case the last row elements are non-zero: $\widetilde{T}_{CC^{*}}(\boldsymbol{f}), \widetilde{T}_{CD^{*}}(\boldsymbol{f}), \widetilde{T}_{DC^{*}}(\boldsymbol{f}), \widetilde{T}_{DD^{*}}(\boldsymbol{f})$. Elements of these matrices can be computed as follows:
\begin{equation}
\widetilde{T}_{XY^{*}}(\boldsymbol{f}) = \Pi^{-1} \int P_{X} (\boldsymbol{x}) P_{Y}^{*} (\boldsymbol{x} + \lambda \boldsymbol{f}) \exp{\bigl\{ -i (\phi(\boldsymbol{x})-\phi(\boldsymbol{x} + \lambda \boldsymbol{f})) \bigr\}} d \boldsymbol{x},
\label{eq:genOTF}
\end{equation}
where $X, Y$ run over $A, B, C, D$, $\phi(\boldsymbol{x})$ --- variations of phase induced by the atmosphere, $\Pi$ --- total area of aperture. For polarization computations a Jones matrix \citep{BornWolf} constitutes analogue of the aperture function:
\begin{equation}
\boldsymbol{P}(\boldsymbol{x}) = \begin{pmatrix}
P_{A}(\boldsymbol{x}) & P_{B}(\boldsymbol{x}) \\
P_{C}(\boldsymbol{x}) & P_{D}(\boldsymbol{x}) \\
\end{pmatrix},
\label{Jones_matrix}
\end{equation}
We use elements of this matrix in formula (\ref{eq:genOTF}). The Jones matrix defines the relation between Jones vectors before ($\boldsymbol{J}$) and after ($\boldsymbol{J}^{\prime}$) optical element:
\begin{equation}
\begin{pmatrix}
J_{x}^{\prime}(\boldsymbol{x}) \\
J_{y}^{\prime}(\boldsymbol{x}) \\
\end{pmatrix} = \begin{pmatrix}
P_{A}(\boldsymbol{x}) & P_{B}(\boldsymbol{x}) \\
P_{C}(\boldsymbol{x}) & P_{D}(\boldsymbol{x}) \\
\end{pmatrix} \begin{pmatrix}
J_{x}(\boldsymbol{x}) \\
J_{y}(\boldsymbol{x}) \\
\end{pmatrix}.
\end{equation}
Outside the aperture all elements of this matrix equal zero.

\begin{figure}[t!]
\begin{center}
\epsfxsize=16cm
\epsffile{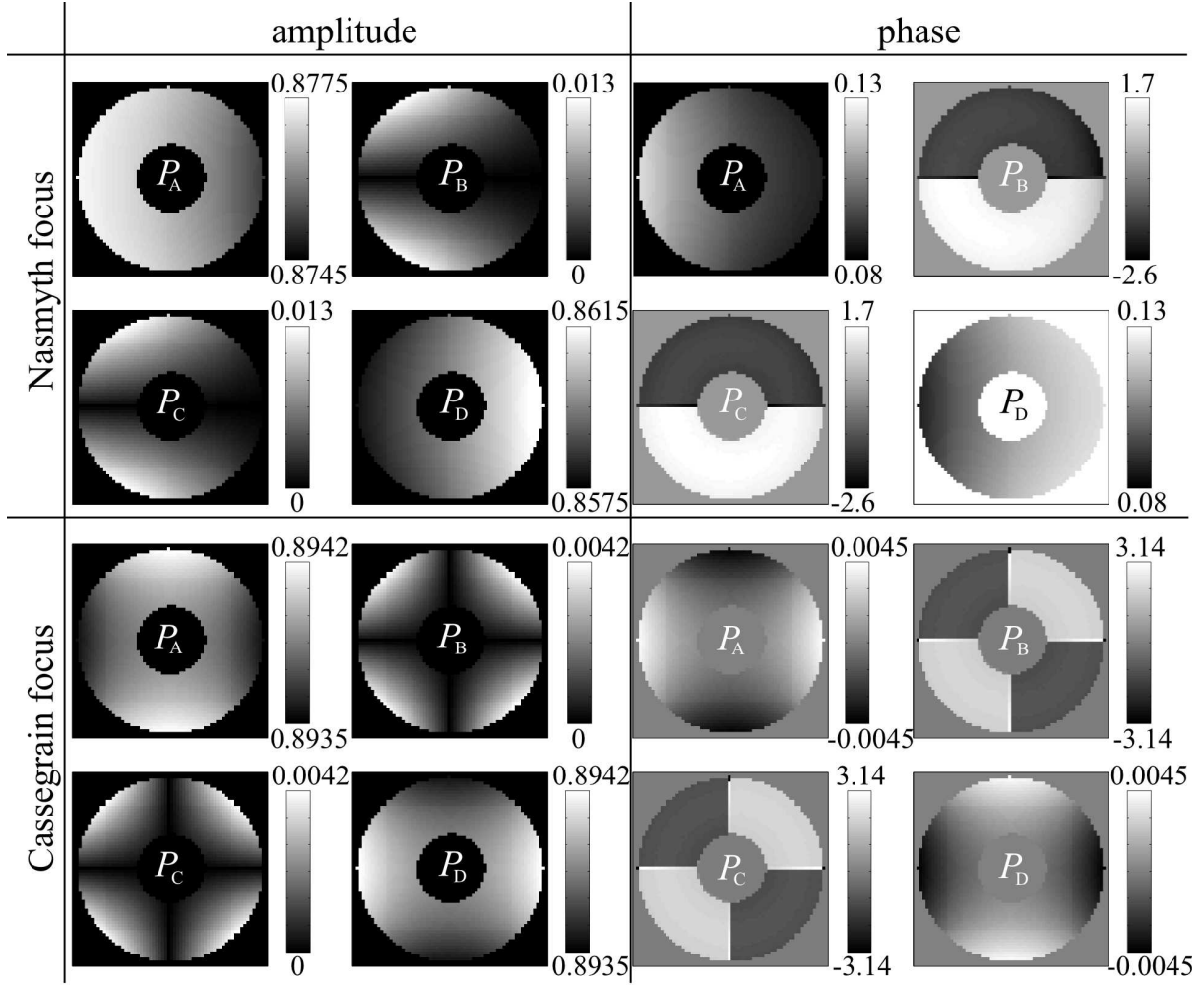}
\end{center}
\caption{Jones matrices. First and second row --- computation for the Nasmyth focus, third and fourth --- for the Cassegrain focus. First and second column --- amplitudes (from left to right, from top to bottom: $P_{A}, P_{B}, P_{C}, P_{D}$), third and fourth --- phases (from left to right, from top to bottom: $P_{A}, P_{B}, P_{C}, P_{D}$).\label{fig:jones}}
\end{figure}

Amplitudes and phases of the Jones matrix for Nasmyth focus of the 2.5-m telescope are plotted in Fig.~\ref{fig:jones}. The computation has been performed with {\it Zemax} software. Evaluation shows that the amplitude of diagonal elements $P_A, P_D$ has mean over the pupil $\approx0.87$ and fluctuation $\approx0.2-0.3\%$. In contrast to them non-diagonal elements have small mean $1\%$ and fluctuations of the same order $1\%$. Thus the amplitude is much smaller for non-diagonal elements then for diagonal. Phases of the diagonal elements differs by $\approx0.1$~rad. Results for the Cassegrain focus are also given in Fig.~\ref{fig:jones}. Absence of oblique reflection gives rise to much smaller variations of the $\boldsymbol{P}$ matrix elements. For diagonal elements $P_A, P_D$ variations of the amplitude amount to $0.05\%$, for non-diagonal --- $0.2\%$. Difference in phases of diagonal elements also is much smaller --- $0.01$~rad.


Inserting all elements of (\ref{eq:genOTF}) into equation (\ref{eq:spec_Mueller}) and using expression (\ref{eq:focal_Mueller_prod}) we obtain Fourier spectrum of Stokes parameters distribution in the focal plane $\widetilde{\boldsymbol{S}}_{f(h,v)}(\boldsymbol{f})$. Taking into account that we measure the intensity only, we obtain resulting expressions for its Fourier spectrum:
\begin{equation}
\begin{aligned}
\widetilde{G}_h & = (\widetilde{T}_{AA^{*}}+\widetilde{T}_{BB^{*}})\widetilde{I}+(\widetilde{T}_{AA^{*}}-\widetilde{T}_{BB^{*}})\widetilde{Q}+(\widetilde{T}_{AB^{*}}+\widetilde{T}_{BA^{*}})\widetilde{U}+i(\widetilde{T}_{BA^{*}}-\widetilde{T}_{AB^{*}})\widetilde{V}, \\
\widetilde{G}_v & = (\widetilde{T}_{CC^{*}}+\widetilde{T}_{DD^{*}})\widetilde{I}+(\widetilde{T}_{CC^{*}}-\widetilde{T}_{DD^{*}})\widetilde{Q}+(\widetilde{T}_{CD^{*}}+\widetilde{T}_{DC^{*}})\widetilde{U}+i(\widetilde{T}_{DC^{*}}-\widetilde{T}_{CD^{*}})\widetilde{V}. \\
\end{aligned}
\label{eq:images}
\end{equation}
Here the $\widetilde{I}, \widetilde{Q}, \widetilde{U}, \widetilde{V}$ --- Fourier spectra of Stokes parameters distribution of the object. As long as evaluation shows that non-diagonal elements $P_{B},P_{C}$ of the Jones matrix have first order of vanishing relatively to diagonal $P_{A},P_{D}$, equations (\ref{eq:images}) can be simplified given that fraction of polarized flux is small $I\gg Q,U,V$:
\begin{equation}
\begin{aligned}
\widetilde{G}_h(\boldsymbol{f}) & \approx \widetilde{T}_{AA^{*}}(\boldsymbol{f})(\widetilde{I}(\boldsymbol{f})+\widetilde{Q}(\boldsymbol{f})), \\
\widetilde{G}_v(\boldsymbol{f}) & \approx \widetilde{T}_{DD^{*}}(\boldsymbol{f})(\widetilde{I}(\boldsymbol{f})-\widetilde{Q}(\boldsymbol{f})). \\
\end{aligned}
\label{eq:images1_A}
\end{equation}

After substitution of these equations into (\ref{eq:phot_add}) and taking into account unknown phase factors we obtain Fourier spectra of detected images
\begin{equation}
\begin{aligned}
\widetilde{F}_h(\boldsymbol{f}) & = \bigl[\widetilde{T}_{AA^{*}}(\boldsymbol{f})(\widetilde{I}(\boldsymbol{f})+\widetilde{Q}(\boldsymbol{f})) + \eta_h(\boldsymbol{f})\bigr] e^{i\pi(\boldsymbol{\theta}_h\cdot\boldsymbol{f})}, \\
\widetilde{F}_v(\boldsymbol{f}) & = \bigl[\widetilde{T}_{DD^{*}}(\boldsymbol{f})(\widetilde{I}(\boldsymbol{f})-\widetilde{Q}(\boldsymbol{f})) + \eta_v(\boldsymbol{f})\bigr] e^{i\pi(\boldsymbol{\theta}_v\cdot\boldsymbol{f})}. \\
\end{aligned}
\label{eq:images1} 
\end{equation}
Substitution of (\ref{eq:images1}) into (\ref{frac_def}) gives in turn:
\begin{equation}
\mathcal{R} = \frac{\left\langle \bigl[ \widetilde{T}_{AA^{*}}(\widetilde{I}+\widetilde{Q}) + \eta_h \bigr] \bigl[ \widetilde{T}_{DD^{*}}(\widetilde{I}-\widetilde{Q}) + \eta_v \bigr]^* \right\rangle_{M}e^{i\pi((\boldsymbol{\theta}_h-\boldsymbol{\theta}_v)\cdot\boldsymbol{f})}}{\left\langle \bigl[ \widetilde{T}_{DD^{*}}(\widetilde{I}-\widetilde{Q}) + \eta_v \bigr] \bigl[ \widetilde{T}_{DD^{*}}(\widetilde{I}-\widetilde{Q}) + \eta_v \bigr]^* \right\rangle_{M}-K_v^{-1}},
\label{eq:RM_gen0}
\end{equation}
where dependence on $\boldsymbol{f}$ is omitted for brevity.


{\color{black}
\section*{Appendix B. Evaluation of $\mathcal{R}$ mean}
\label{subs:Rbias}

The value $\mathcal{R}$ represents ratio of sample means computed from $M$ measurements. From statistics it is known that sample mean of some value is a normally distributed random number with mean equal to mean of this value. Therefore, given that $M$ is sufficiently large, $\mathcal{R}$ is also normally distributed random number with mean equal to the ratio of means of values $\widetilde{F}_h \widetilde{F}_v^*$ and $\widetilde{F}_v \widetilde{F}_v^* - K_v^{-1}$. Let us denote this mean as $\overline{\mathcal{R}}$ and let us estimate it. The equation (\ref{eq:RM_gen0}) is taken as a starting point.


Let us introduce the following notations:
\begin{equation}
N_h(\boldsymbol{f}) = \frac{\eta_h(\boldsymbol{f})}{\widetilde{I}(\boldsymbol{f})+\widetilde{Q}(\boldsymbol{f})}, ~~~~ N_v = \frac{\eta_v(\boldsymbol{f})}{\widetilde{I}(\boldsymbol{f})-\widetilde{Q}(\boldsymbol{f})},
\end{equation}
where $\widetilde{I}(\boldsymbol{f})+\widetilde{Q}(\boldsymbol{f})$ and $\widetilde{I}(\boldsymbol{f})-\widetilde{Q}(\boldsymbol{f})$ are constant in time.


Given that fluctuations of $P_{A}(\boldsymbol{x})$ and $P_{D}(\boldsymbol{x})$ are small is comparison with the values themselves the $\widetilde{T}_{AA^{*}}$ and $\widetilde{T}_{DD^{*}}$ can be transformed (we perform derivations for $P_{A}(\boldsymbol{x})$, for $P_{D}(\boldsymbol{x})$ they are analogous):
\begin{equation}
P_{A}(\boldsymbol{x}) = P_0(\boldsymbol{x})+\Delta P_{A}(\boldsymbol{x}),
\end{equation}
where $P_0(\boldsymbol{x})$ is the aperture function in the usual sense, it equals one inside pupil and zero outside, $\Delta P_{A}(\boldsymbol{x}) \ll 1$. Inserting this into integrand of (\ref{eq:genOTF}) we get (neglecting the cross term, because it has second order of vanishing):
\begin{equation}
\widetilde{T}_{AA^{*}}(\boldsymbol{f}) = \widetilde{T}_{\mbox{atm}}(\boldsymbol{f}) + \Delta \widetilde{T}_{AA^{*}}(\boldsymbol{f}),
\label{eq:OTF_approx}
\end{equation}
where $\widetilde{T}_{\mbox{atm}}(\boldsymbol{f})$  is OTF of system telescope + atmosphere in the usual sense:
\begin{equation}
\widetilde{T}_{\mbox{atm}}(\boldsymbol{f}) = \Pi^{-1} \int P_0(\boldsymbol{x}) P_0^*(\boldsymbol{x}+\lambda\boldsymbol{f}) \exp{\bigl\{ -i (\phi(\boldsymbol{x})-\phi(\boldsymbol{x} + \lambda \boldsymbol{f})) \bigr\}} d \boldsymbol{x}.
\end{equation}
$\Delta \widetilde{T}_{AA^{*}}(\boldsymbol{f})$ is some correction factor similar to OTF:
\begin{equation}
\begin{split}
\Delta \widetilde{T}_{AA^{*}}(\boldsymbol{f}) = \Pi^{-1} \int & P_0(\boldsymbol{x}) P_0^*(\boldsymbol{x}+\lambda\boldsymbol{f}) \bigl[\Delta P_{A}(\boldsymbol{x}) + \Delta P_{A}^{*}(\boldsymbol{x}+\lambda \boldsymbol{f}) \bigr] \\ 
& \times \exp{\bigl\{ -i (\phi(\boldsymbol{x})-\phi(\boldsymbol{x} + \lambda \boldsymbol{f})) \bigr\}} d \boldsymbol{x}.\\
\label{eq:OTF_corr}
\end{split}
\end{equation}
$\Pi$ is a total aperture area. Substituting (\ref{eq:OTF_approx}) into (\ref{eq:RM_gen0}) and after development we obtain
\begin{equation}
\overline{\mathcal{R}} = \mathcal{R}_0 \frac{\Bigl\langle \bigl[ \widetilde{T}_{\mbox{atm}}  + \Delta \widetilde{T}_{AA^{*}}  + N_h  \bigr] \bigl[ \widetilde{T}_{\mbox{atm}}  + \Delta \widetilde{T}_{DD^{*}}  + N_v  \bigr]^* \Bigr\rangle}{\Bigl\langle \bigl[ \widetilde{T}_{\mbox{atm}}  + \Delta \widetilde{T}_{DD^{*}}  + N_v  \bigr] \bigl[ \widetilde{T}_{\mbox{atm}}  + \Delta \widetilde{T}_{DD^{*}}  + N_v  \bigr]^* \Bigr\rangle-K_v^{-1}},
\label{eq:RM_gen2}
\end{equation}
where dependence on $\boldsymbol{f}$ is omitted for brevity. Hereafter angle parenthesis means averaging over assembly.

Now we are going to recast (\ref{eq:RM_gen2}): remove parenthesis, take into account that $\Delta \widetilde{T}_{AA^{*}}, \Delta \widetilde{T}_{DD^{*}} \ll \widetilde{T}_{\mbox{atm}}$, $\langle N_h \rangle = \langle N_v \rangle = 0$ (as long as they are circularly symmetric complex normal random numbers) and $\langle \widetilde{T}_{\mbox{atm}} \rangle = \langle \Delta \widetilde{T}_{AA^{*}} \rangle = \langle \Delta \widetilde{T}_{DD^{*}} \rangle = 0$ for frequencies $f > r_0/\lambda$ (absence of high spatial frequencies in long exposure image):
\begin{equation}
\overline{\mathcal{R}}(\boldsymbol{f}) = \mathcal{R}_0 \Biggl( 1 + \frac{\langle \widetilde{T}_{\mbox{atm}}^* \Delta \widetilde{T} \rangle}{\langle \widetilde{T}_{\mbox{atm}} \widetilde{T}_{\mbox{atm}}^*\rangle} \Biggr),
\label{eq:RM_gen3}
\end{equation}
where $\Delta \widetilde{T} = \Delta \widetilde{T}_{AA^{*}} - \Delta \widetilde{T}_{DD^{*}}$. Let us develop it explicitly:
\begin{equation}
\Delta \widetilde{T}(\boldsymbol{f}) = \Pi^{-1} \int P_0(\boldsymbol{x}) P_0^*(\boldsymbol{x}+\lambda\boldsymbol{f}) \Delta P(\boldsymbol{x},\boldsymbol{f}) \exp{\bigl\{ -i (\phi(\boldsymbol{x})-\phi(\boldsymbol{x} + \lambda \boldsymbol{f})) \bigr\}} d\boldsymbol{x},
\label{eq:deltaD_def}
\end{equation}
where $\Delta P$:
\begin{equation}
\Delta P(\boldsymbol{x},\boldsymbol{f}) = \Delta P_A(\boldsymbol{x}) + \Delta P_A^*(\boldsymbol{x}+\lambda \boldsymbol{f}) - \Delta P_D(\boldsymbol{x}) - \Delta P_D^*(\boldsymbol{x}+\lambda \boldsymbol{f}).
\end{equation}


For numerator of the second term in (\ref{eq:RM_gen3}) one can write:
\begin{equation}
\begin{split}
\langle \widetilde{T}_{\mbox{atm}}^* \Delta \widetilde{T} \rangle = & \Pi^{-2} \iint  P_0(\boldsymbol{x}) P_0^*(\boldsymbol{x}+\lambda\boldsymbol{f}) \Delta P(\boldsymbol{x},\boldsymbol{f})  \\
& \times \Bigl\langle \exp{\bigl\{ -i (\phi(\boldsymbol{x})-\phi(\boldsymbol{x} + \lambda \boldsymbol{f}) - \phi(\boldsymbol{x^{\prime}})+\phi(\boldsymbol{x^{\prime}} + \lambda \boldsymbol{f})) \bigr\}} \Bigr\rangle d\boldsymbol{x} d\boldsymbol{x^{\prime}}.
\end{split}
\end{equation}


Averaged expression in this integral equals $\sigma\delta(\boldsymbol{x}-\boldsymbol{x^{\prime}})$ for frequencies $f\gg r_0/\lambda$, where $\sigma$ is atmospheric coherence area, which equals $0.342 r_0^2$ for the Kolmogorov turbulence \citep{Korff1973}. Using this fact we obtain:
\begin{equation}
\langle \widetilde{T}_{\mbox{atm}}^* \Delta \widetilde{T} \rangle = {(L\Pi)}^{-1} \int P_0(\boldsymbol{x}) P_0^*(\boldsymbol{x}+\lambda\boldsymbol{f}) \Delta P(\boldsymbol{x},\boldsymbol{f}) d\boldsymbol{x},
\end{equation}
where $L$ --- total number of speckles in the image.


On the same assumptions one can show that \citep{Korff1973}
\begin{equation}
\langle \widetilde{T}_{\mbox{atm}} \widetilde{T}_{\mbox{atm}}^* \rangle = {(L\Pi)}^{-1} \int P_0(\boldsymbol{x}) P_0^*(\boldsymbol{x}+\lambda\boldsymbol{f}) d\boldsymbol{x},
\end{equation}
where integral is OTF of the system in the absence of the atmosphere $\widetilde{T}_0(\boldsymbol{f})$. Ultimately for $\overline{\mathcal{R}}(\boldsymbol{f})$:
\begin{equation}
\overline{\mathcal{R}}(\boldsymbol{f}) = \mathcal{R}_0(\boldsymbol{f}) \bigl( 1 + \Delta \mathcal{R}(\boldsymbol{f}) \bigr),
\label{eq:RM_gen4}
\end{equation}
where
\begin{equation}
\Delta \mathcal{R}(\boldsymbol{f}) = \int P_0(\boldsymbol{x}) P_0^*(\boldsymbol{x}+\lambda\boldsymbol{f}) \Delta P(\boldsymbol{x},\boldsymbol{f}) d\boldsymbol{x} \Biggr/ \int P_0(\boldsymbol{x}) P_0^*(\boldsymbol{x}+\lambda\boldsymbol{f})  d\boldsymbol{x}.
\label{eq:deltaR}
\end{equation}
Thus $\Delta \mathcal{R}(\boldsymbol{f})$ is $\Delta P(\boldsymbol{x},\boldsymbol{f})$ averaged over the area defined by $P_0(\boldsymbol{x})P_0^*(\boldsymbol{x}+\lambda\boldsymbol{f})$.


\section*{Appendix C. Evaluation of $\mathcal{R}$ variance}
\label{subs:Rdisp}

Here the method of evaluation of variance of $\mathcal{R}$ from simulated data is given. As can be seen from (\ref{frac_def}), $\mathcal{R}$ can be considered as a function of three random numbers $X$, $Y$, $Z$:
\begin{equation}
\mathcal{R} = \frac{X+i Y}{Z-K_v^{-1}},
\label{eq:Rxyz}
\end{equation}
where $X = \mbox{Re} \langle \widetilde{F}_h \widetilde{F}_v^* \rangle_M$, $Y = \mbox{Im} \langle \widetilde{F}_h \widetilde{F}_v^* \rangle_M$, $Z = \langle \widetilde{F}_v \widetilde{F}_v^* \rangle_M$.
%

For further use it is convenient to decompose the variance of the $\mathcal{R}$ into variance $\sigma_A^2$ of its amplitude and variance $\sigma_{\phi}^2$ of its phase. On the assumption that standard deviations of $X$, $Y$, $Z$ are small relatively to the values themselves, one can express $\sigma_A^2$ and $\sigma_{\phi}^2$ as follows:
\begin{equation}
\sigma_A^2 = \sigma_{XX}^2\frac{X^2}{Z^{\prime2} W^2} + \sigma_{YY}^2\frac{Y^2}{Z^{\prime2} W^2} + \sigma_{ZZ}^2\frac{W^2}{Z^{\prime4}} + \sigma_{XY}^2\frac{2 X Y}{Z^{\prime2} W^2} - \sigma_{XZ}^2\frac{2X}{Z^{\prime3}} - \sigma_{YZ}^2\frac{2Y}{Z^{\prime3}},
\label{eq:RnoiseA}
\end{equation}
\begin{equation}
\sigma_{\phi}^2 = \frac{\sigma_{XX}^2 Y^2 + \sigma_{YY}^2 X^2 - 2 \sigma_{XY}^2 X Y}{W^4},
\label{eq:RnoisePhi}
\end{equation}
where $\sigma_{XX}^2, \sigma_{XY}^2$, etc. are elements of the covariance matrix of values $X$, $Y$, $Z$. Here we use delta method and introduce two auxiliary values for convenience: $Z^{\prime} = Z - K_v^{-1}$ and $W = \sqrt{X^2+Y^2}$.

}

\end{document}